\documentclass{pasj00}
\draft

\begin{document}
\SetRunningHead{Liu, Fu, Deng, \& Huang}{The Equilibrium Burkert
Profile}
\Received{2005/04/06}%{yyyy/mm/dd}
\Accepted{2005/05/29}%{yyyy/mm/dd}

\title{An Equilibrium Dark Matter Halo with the Burkert Profile}

%%% begin:list of authors
\author{Wen-Hao \textsc{Liu}}
  %\thanks{Example: Present Address is xxxxxxxxxx}}
\affil{Department of Astronomy, Nanjing University,Nanjing,210093,
China } \email{dracoliu@hotmail.com}

\author{Yan-Ning \textsc{Fu}}
\affil{Purple Mountain Observatory, Chinese Academy of
Sciences,Nanjing 210008,
and\\
           National Observatories, Chinese Academy of Sciences,
           Beijing 100039, China}\email{fyn@pmo.ac.cn}

\author{Zu-Gan \textsc{Deng}}
\affil{Department of Physics, Graduate School, Chinese Academy of
Sciences, Beijing 100039, China}\email{dzg@vega.pku.bac.ac.cn}
 \and
\author{Jie-Hao {\sc Huang}}
\affil{Department of Astronomy, Nanjing University, Nanjing
210093, China} \email{jhh@nju.edu.cn}
%%% end:list of authors

%%% Please use the following style in case that sorting by
%%% affilation is impossible.
%
% \author{%
%   D-Firstname \textsc{D-Familyname}\altaffilmark{1}
%   E-Firstname \textsc{E-Familyname}\altaffilmark{1,2}
%   and
%   F-Firstname \textsc{F-Familyname}\altaffilmark{2}}
% \altaffiltext{1}{Address of Institute}
% \email{ddddd@xxx.xxx.xx.xx}
% \email{eeeee@xxx.xxx.xx.xx}
% \altaffiltext{2}{Address of Institute}

%% `\KeyWords{}' always has to be placed before `\maketitle'.
\KeyWords{cosmology: dark matter - galaxies: kinematics and
dynamics - methods: numerical}
%Do NOT move this preamble from here!

\maketitle

\begin{abstract}
In this paper a prescription for generating an equilibrium
spherical system depicted with cuspy density profiles is extended
to a core density profile, the Burkert profile, which is
observationally more suitable to dwarfs. By using a time-saving
Monte Carlo method instead of N-body simulations, we show that the
Burkert halo initialized with the distribution function that
depends only on energy is in a practically stable equilibrium. The
one generated using the local Maxwellian approximation is
unstable, where the flat core density structure tends to steepen.
This is fundamentally different from the previous study on a halo
with cuspy density profile. The deviation of the unstable
"Burkert" from the initial Burkert profile is found to be closely
related to taking a Gaussian as the velocity distribution at any
given point. The significance of not using the local Maxwellian
approximation is further demonstrated by exploring the dynamical
evolution of compact super star clusters in the Burkert halo. In
particular, the local Maxwellian approximation will result in
underestimating the dynamical friction and overestimating sinking
time scale. Accordingly, it will lead to lower probability of
forming massive bulges and young nuclear star clusters in
bulgeless galaxies.
\end{abstract}

\section{Introduction}

%Dynamical Studies of galaxy evolution and formation commonly
%consider the self-consistent behavior of one component or the
%interaction between two components; e.g., the effects of tidal
%stripping and dynamical friction on the orbits of satellites
%(\cite{vel95}; \cite{may02}), bulge formation in late-type
%galaxies (\cite{fu03a}), globular cluster formation with dark
%matter halos (\cite{mas04}), the inner structure of $\Lambda$CDM
%halos (\cite{nav04}). It is more appropriate to choose an isolated
%equilibrium galaxy or halo model and use high-resolution
%simulations to investigate such problems.

Generating initial conditions for numerical experiments of a system
with a certain density profile is a well-defined but difficult
procedure. There are two steps for constructing the initial
conditions of one desired model: 1) calculate the steady-state
distribution function of the desired model; 2) use Monte Carlo
method to generate the initial conditions. The main difficulty comes
from the first step. One possibility is to make some simplifying
approximations about the nature of the distribution function
(\cite{tom72}; \cite{her88}). Hernquist (1993) described a
prescription for constructing N-body realization assuming that
velocity distribution at any given point is Maxwellian. It is called
the local Maxwellian approximation (\cite{kaz04}), of which the
advantage is easy to implement. However, Kazantzidis, Magorrian \&
Moore (2004, KMM hereafter) show that the halo initialized using the
local Maxwellian approximation (the Maxwellian halo hereafter for
simplicity ) can be significantly far from equilibrium, and that for
high resolution simulations it is more reasonable to use a stable
system generated with a certain steady-state distribution function
(the DF halo hereafter).

In particular, they demonstrated that an unstable cuspy halo
constructed using the local Maxwellian approximation will soon lose
its cusp structure due to its own evolution, which is previously
claimed to be due to minor merger between a host halo and the
falling sub-system. On the contrary, a stable cuspy halo survives
the minor merger. This investigation indicates that the
non-equilibrium Maxwellian halo will surely lead to spurious
evolution of the host halo when dealing with the response, e.g. the
responding of a host halo to sinking SSCs will otherwise be mixed
with the artificial evolution of the host halo, induced by its
unstability. This effect has been investigated in our recent work on
the bulge formation in late-type spirals (\cite{fu05}).

The halos on which KMM studied are those with cuspy structure.
They have not explored the halos with flat inner profile, e.g. the
Burkert profile. The progress in observations (e.g. \cite{mar02};
and the references therein) shows, however, that the cuspy DM
profiles, such as the NFW (\cite{nav97}) profile, are not adequate
for a large fraction of dwarf galaxies, which are dominated by DM
halos, and suggests that the core density profiles, the Burkert
(1995) profile for example, are more suitable to these galaxies.

It would then be very much instructive to extend KMM's study to the
Burkert halo. It is the halo that contains less mass in the central
region than those in the cuspy halos, e.g. the NFW profile. It
follows that the technical difficulties will arise for
high-resolution simulations of the core halos, especially for those
studies focused on the central regions (\cite{fu03a};~\cite{hua03}).
For a given mass of DM halo, our test shows that in order to reach
the same resolution in the central 10$pc$ region the number of
particle for the Burkert halo should be a hundred times more than
that for the NFW halo. It would need much more computing time with
the N-body simulations.

In this paper, we aim at describing the way to generate the
initial conditions for an isotropic, spherical Burkert halo, and
perform simulations using the Monte Carlo method (Henon 1971a,b)
to study the stable situation of the Maxwellian and the DF Burkert
halos. This may have important implications for the dynamical
evolution of SSCs in a responding dark matter halo, resulting in
different bulge formation history in late-type spirals.

The organization of the paper is as follows. The procedure to
constructing the initial conditions for our model is described in
section 2. Section 3 gives the results of the simulations. We
discuss the results and illustrate the important implications of
generating the DF halo in Section 4. Finally, Section 5 notes our
conclusion.

%\newpage

\section{Models and Methods}

\subsection{Density Profile}
Here we study the Burkert dark matter halo (1995):
\begin{equation}
  \rho=\frac{\rho_{s}}{(1+r/r_{s})[1+(r/r_{s})^{2}]}
\end{equation}
where the central density $\rho_{s}$ is taken to be
0.05$M_{\odot}/pc^{3}$ following Marchesini et al. (2002), and the
scale radius, $r_{s}$ is computed from the mass of the DM halo
$M_{200}=10^{11}M_{\odot}$. In the inner part of the Bukert halo
the profile has a core structure, while the slope approximates to
$-3$ in the infinity. The cumulative mass profile with such a
distribution diverges as $r\rightarrow\infty$. In fact, the
Burkert profile is a phenomenological formula, which provides a
good fit to the observed data to less than the virial radius
$r_{vir}$ and not valid to arbitrarily large distance from the
galactic center. Instead of truncating the profile sharply at
$r_{vir}$, we have chosen an exponential cut-off for $r>r_{vir}$
following KMM, which sets in at $r_{vir}$ and turn-off in a scale
$r_{decay}$.
\begin{equation}
\rho=\frac{\rho_{s}}{(1+c)(1+c^{2})}(\frac{r}{r_{vir}})^{\alpha}~exp[-\frac{r-r_{vir}}{r_{decay}}]~~~~~~~~~for
  (r>r_{vir})
\end{equation}
where $c\equiv{r_{vir}}/{r_{s}}$ is the concentration parameter.
In order to ensure a smooth transition between eq.(1) and eq.(2)
at $r_{vir}$, we require the logarithmic slope there to be
continuous. This leads to
\begin{equation}
  \alpha=\frac{3r_{vir}}{r_{decay}}-\frac{2c^{2}}{1+c^{2}}-\frac{c}{1+c}
\end{equation}
We consider the simulations between a minimum radius $r_{min}$ and
a maximum radius $r_{max}$. The minimum radius is chosen such that
it has sufficient particles for us to analyze; The maximum radius
is equal to the virial radius plus several $r_{decay}$. Here we
adopt three $r_{decay}$ ($r_{decay}=0.1r_{vir}$). In addition, we
add a boundary at the maximum radius. If the halo is stable, it is
in homeostasis, that is, at any time the number of particles which
move out off the boundary is equal to that of particles which move
in. In the case of the Burkert halo, the DF (Burkert) halo will
not be stable in the outer region, if not adding such a boundary.
This is because, not like the case of KMM, the mass outside
$r_{vir+3rdecay}$ is still not negligible. In other words, a sharp
truncation there will make the DF halo unstable.

\subsection{Distribution Function}
In this paper, we restrict our models to isotropic, non-rotating
halos. According to the Jeans$'$ theorem (\cite{lyn62};
\cite{bin87}), the distribution function of any steady-state
spherical system can be expressed as $f(E,\mathbf{L})$, where E is
the binding energy and $\mathbf{L}$ is the angular momentum
vector. As usual, the relative potential $\Psi$ and the relative
energy $\varepsilon$ (a dimensionless energy in units of
$\pi^{2}G\rho_{s}r_{s}^{2}$) are defined as:
\begin{equation}
\Psi\equiv -\Phi+\Phi_{0}; \varepsilon\equiv -E+\Phi_{0}
\end{equation}
where $\Phi$ is the potential and $\Phi_{0}$ is a constant
fulfilling $f>0$ for $\varepsilon>0$ and $f=0$ for
$\varepsilon\leq0$. In an isotropic spherical model the
distribution function depends on $\varepsilon$ only, that is,
$f=f(\varepsilon)$. Integrating $f$ over all velocities, we can
get the density profile:
\begin{equation}
\rho(r)=\int{f(\varepsilon)}d^{3}\mathbf{v}=4\pi\int_{0}^{\Psi}{f(\varepsilon)\sqrt{2(\Psi-\varepsilon)}}d\varepsilon
\end{equation}
The inversion of the above equation gives the distribution
function (\cite{edd16}; \cite{bin87}),
\begin{equation}
f(\varepsilon)=\frac{1}{\sqrt{8}\pi^{2}}[\int_{0}^{\varepsilon}\frac{d^{2}\rho}{d\Psi^{2}}\frac{d\Psi}{(\sqrt{\varepsilon-\Psi})}+\frac{1}{\sqrt{\varepsilon}}(\frac{d\rho}{d\Psi})_{\Psi=0}]
\end{equation}
The $d^{2}\rho/d\Psi^{2}$ factor can be evaluated from eq.(1) and
eq.(2). The second term of the right-hand side vanishes in large
distance in our model.

\subsection{Initialization Procedure}
Here we describe two approaches to initialize the numerical
simulations for the Burkert profile using the local Maxwellian
approximation and the steady-state distribution function,
respectively.

\subsubsection{Exact Distribution Function}
Having the density profile $\rho(r)$, we can calculate the model's
cumulative mass distribution M(r) and the gravitational potential
$\Phi(r)$. Then we can calculate the distribution function from
eq.(6).

The integrand in eq.(6) diverges at one or both of the limits, but
this can be solved using standard techniques for improper integral
(\cite{pre86}). The distribution function is obtained on a grid of
$\varepsilon$ equally spaced in $\log(\varepsilon)$. The accuracy
of the numerical integration is checked by comparing the density
profile derived from eq.(5) and that given by eq.(1). The two
profiles are very well accordant. We find that the distribution
function is non-negative everywhere, which proves that the
assumption of the isotropic velocity distribution for the
spherical Burkert model is reasonable and physical.

The procedure to obtain the physical quantities of the sampling
particles is the following. We randomly sample the initial
positions and velocities of the particles. For our model, the
probability density function (Windrow 2000) of the particle having
the relative energy $\varepsilon$ at the radius R is:
\begin{equation}
  P(\varepsilon,R)\propto R^{2}(\Psi(R)-\varepsilon)^{1/2}f(\varepsilon)
\end{equation}

We use the rejection method (\cite{pre86}) to sample the particles
in two steps: 1) Generate the position of the particle. The
probability density function with radius R is:
\begin{equation}
  P(R)\propto R^{2}\rho(R)
\end{equation}
2) Generate the energy of the particle. Once given R, we can get
the relative potential $\Psi(R)$. Then the probability density
function having the relative energy $\varepsilon$ and potential
$\Psi(R)$ is:
\begin{equation}
  P(\varepsilon)\propto (\Psi(R)-\varepsilon)^{1/2}f(\varepsilon)
\end{equation}
Once the position R and the relative energy $\varepsilon$ of a
particle are given, the speed v can be easily determined by
$\varepsilon$=$v^{2}/2$+$\Psi$. Finally, the tangential and radial
velocities of each particle can be randomly sampled considering
that our model is isotropic.

\subsubsection{Local Maxwellian Approximation}
The way to sample the position of a particle is the same as that
depicted above. The difference is how to generate the velocity of
the particle. The collisionless Boltzmann equation for a spherical
system can be written as:
\begin{equation}
\frac{d(\rho\overline{v_{r}^{2}})}{dr}+\frac{\rho}{r}[2\overline{v_{r}^{2}}-(\overline{v_{\theta}^{2}}+\overline{v_{\phi}^{2}}]=-\rho\frac{d\Phi}{dr}
\end{equation}
where $\overline{v_{r}^{2}}$, $\overline{v_{\theta}^{2}}$, and
$\overline{v_{\phi}^{2}}$ are the velocity dispersions in
spherical coordinates (\cite{bin87}). Assuming the model is
isotropic, it can be integrated to:
\begin{equation}
\overline{v_{r}^{2}}=\frac{1}{\rho(r)}\int_{r}^{\infty}\rho(r)\frac{d\Phi}{dr}dr
\end{equation}
Using the known density profile, the radial dispersion can be
computed as a function of radius. In the local Maxwellian
approximation, the 1D velocity distribution at a given point is
approximated by a Gaussian (\cite{her93}):
\begin{equation}
F(v,r)=4\pi(\frac{1}{2\pi\sigma^{2}})^{3/2}v^{2}exp(-v^{2}/2\overline{v_{r}^{2}})
\end{equation}

The speed of a halo particle is initialized from the eq.(12). Then
the tangential and radial velocities are randomly sampled in the
same way as that described in the above subsection.

\subsection{Simulations and Parameters}
Many approaches have been used in simulations to study the
galactic dynamics. One is to give the statistical description of
the system, e.g. distribution function
$f(\overrightarrow{r},\overrightarrow{v},t)$. Rosenbluth et al.
(1957) discussed a way to calculate $f$ by solving a much
complicated equation, Fokker-Planck equation. But it needs to make
a number of arbitrary simplifications. A commonly used alternative
way is N-body simulations. Unfortunately, this approach requires
much computing time, which will greatly increase with the number
of particles. As we mentioned before, in the case of a core halo
like the Burkert one (1995), a large number of particles are in
demand to keep high resolution in the central region. In view of
this, we adopt a time-saving method, the Monte Carlo procedure
(Henon 1971a,b), rather than the N-body simulations. This
procedure has been proved to reproduce the behavior of the system
given by the Fokker-Planck equation and to be much faster than the
N-body simulations.

The basic ideas of the Monte Carlo procedure is the following. We
can divide the gravitation filed of the system into two parts: a
main smoothed-out field and a small fluctuating one. The particle
moves during every time step $\triangle$t which satisfies:
\begin{equation}
  T_{c}\ll\triangle t\ll T_{r}
\end{equation}
where $T_{c}$ and $T_{r}$ are the crossing time and the relaxation
time, respectively. The halo evolves accordingly. The relaxation
time is related to the crossing time (\cite{bin87}):
\begin{equation}
  T_{r}/T_{c}\propto N/\ln N
\end{equation}
And the crossing time can be calculated as a function of the total
mass M and the total energy $\varepsilon$ of the system
(\cite{van68}):
\begin{equation}
  T_{c}=CGM^{5/2}|\varepsilon|^{-3/2}
\end{equation}
where C is a numerical constant.

Neglecting the fluctuating field in a first approximation, the
motion of the particle is governed by the main spherical symmetry
field which changes with the density or mass distribution of the
halo. The particle will then take a plane rosette motion
(\cite{bin87}). In any given state the particle is characterized by
its position, energy and its angular momentum.  One could derive the
statistic of the system by randomly choosing the new positions and
velocities of the particles on their respective orbits. Taking them
as the new conditions, we can calculate a new density (or mass)
distribution and accordingly a new potential of the system, which
will in turn determine the positions of particles at the next step.
The equilibrium state of the halo can be tested by comparing the new
density (or mass) distribution with the initial one. Due to this
character, if we find some deviation from the initial state of a
system at one step, the deviation should grow in the simulations
until the system relaxes to a new stable equilibrium (as
Figure~\ref{DenProvsD}b shows). In other words, what we could
illustrate with the Monte Carlo method  at the present time is that
the system is practically stable or unstable.

For our simulations, the relaxation time of the system is much
long and we need not consider the two-body relaxation in our
study. If the system is in stable equilibrium, the statistic of
the system such as the potential field (or the density and mass
distribution) will keep the same though the positions and
velocities of the particles are randomly chosen in each step. On
the contrary, in an unstable system, from the density or mass
profile we can know that the statistic of the system have changed
after one step.

The parameters of our simulations are given in Table 1.
\begin{table}[]
  \caption[1]{Parameters of Simulations}
%% \label{Tab:parameter}
  \begin{center}\begin{tabular}{clclclclcl}
  \hline\hline\noalign{\smallskip}
Model&  $M_{total}$  &   $\rho_{s}$      & $r_{s}$ & $r_{vir}$ &
$r_{max}$ & N  &  initializing method         \\
          &  $M_{\odot}$  &$M_{\odot}/pc^{3}$  &  kpc  &  kpc     &  kpc
&         \\
  \hline\noalign{\smallskip}
A &  $10^{11}$    & 0.05              & 4.0   &  91.24   &  118.2
& 2 $\times
10^{6}$ & exact distribution function\\
B&  $10^{11}$    & 0.05              & 4.0   &  91.24   &  118.2 &
2 $\times
10^{6}$ & local Maxwellian approximation \\
  \noalign{\smallskip}\hline
  \end{tabular}\end{center}
\end{table}

\section{Results}
\begin{figure*}
    \begin{center}\vspace*{-46mm}
    \FigureFile(10.8cm,9.3cm){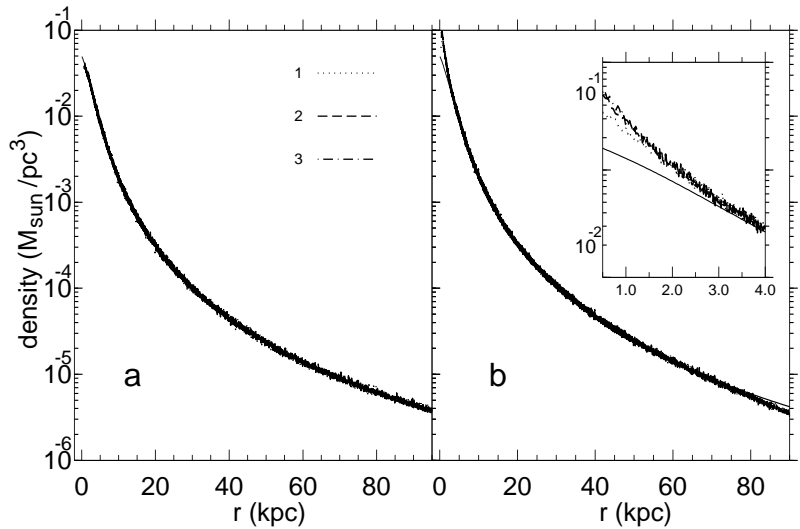}
    \end{center}\vspace*{-21mm}
    \caption{\label{DenProvsD}
Radial density profiles for Model A and B are presented in a) and
b) panel, respectively.
 The dotted, dashed and dot-dashed lines in both panels correspond to the density
 distributions derived at three randomly choosing steps.
 The thin solid lines in two panels show the initial Burkert profile given by eq.(1).
 The insert diagram in b) illustrates the situation in the central region for clarity.}
\end{figure*}

We run our models for 20 steps and generate 20 individual
snapshots. The radial density profiles shown in this section are
obtained by binning the particles from individual snapshot.
Figure~\ref{DenProvsD}a) illustrate the radial density profiles
for Model A, where the density distributions at three different
steps, denoted with dotted, dashed and dot-dashed lines, are
plotted. For comparison, the (initial) Burkert profile given by
eq.(1) is also shown in the figure, denoted with thin solid line.
Obviously the radial density profiles for Model A have no
evolution over the simulations, all of them are indeed very well
matched to the (initial) Burkert profile. In view of this, we
conclude that the DF Burkert halo, i.e. the Burkert halo
initialized with the exact distribution function, is in stable
equilibrium.

The opposite situation  occurs for Model B. We present the results
in Figure~\ref{DenProvsD}b), where the density distributions
derived at three randomly choosing steps are indicated. The
profiles denoted with dotted, dashed and dot-dashed lines
illustrate significant deviation from the (initial) Burkert
profile at the central and outer regions. The central distribution
tends to steepen as clearly shown in the insert diagram of
Figure~\ref{DenProvsD}b) and the outer densities decline. The
disparity from the initial one simply shows that the Maxwellian
Burkert halo, i.e. the Burkert halo generated with the local
Maxwellian approximation is not in equilibrium.

\section{Discussions}
\begin{figure}
   \centering
   \FigureFile(7.8cm,7.3cm){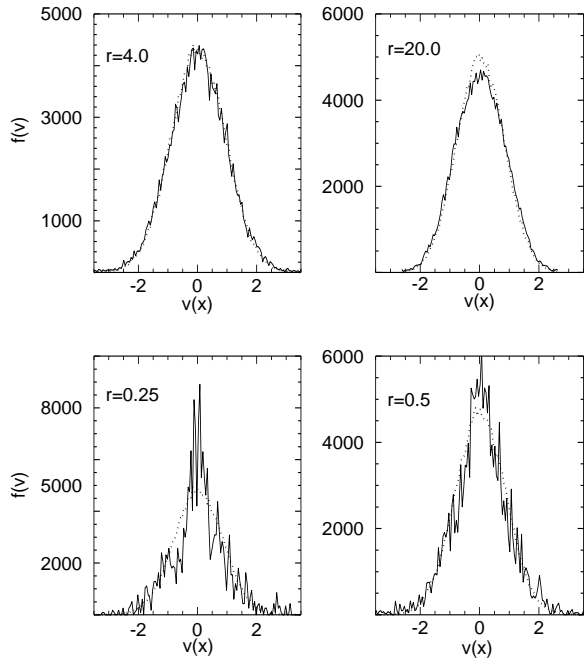}\vspace*{-6mm}
      \caption{Histograms of the one dimensional velocity distribution at four
      different distances from the center in units of $kpc$. The solid lines
correspond to the true velocity distribution, and the dotted lines
show the Gaussian
     velocity profile.
              }
         \label{VelDisvsDisCen}
   \end{figure}

Following KMM, we analyze the actual velocity structure at various
distances for Model A, compared with the corresponding Gaussian
distributions used in the Maxwellian approximation (Model B). The
solid and dotted lines in Figure~\ref{VelDisvsDisCen} correspond
to the true and Gaussian velocity distributions, respectively, at
four different distances from the center. It is evident that only
at about the scale radius the Gaussian distribution is a close
approximation to the reality. The disparity in the velocity
structure from the Gaussian distribution is very strong near the
center. The true velocity distributions there are strongly peaked
than a Gaussian. It is still apparent at far from the center,
where the true velocity structure is shallower than a Gaussian on
the contrary.

The feature in velocity distributions described above is indeed
reflected in the Maxwellian Burkert halo, where the deviation from
the (initial) Burkert halo is obvious at the central and outer
regions. The calculations show that the total energy and angular
momentum of particles in the Maxwellian Burkert halo are less than
those in the DF Burkert halo, i.e. the (initial) Burkert halo.
That is, the Maxwellian Burkert halo is colder than the DF Burkert
halo. This explains the hoist of the central density profile in
the Maxwellian halo. In other words, the unstable situation of the
Maxwellian halo is closely related to taking a Gaussian as the
velocity distribution at any given point.

In some existing N-body simulations, the considered system with a
predetermined mass density profile are initialized using the
Maxwellian approximation. In order to start simulations with a
system in stable or quasi-stable equilibrium, the authors allow
the constructed system to relax for some time so that a ceratin
quasi-stable equilibrium is attained. According to KMM's
investigation and the study presented in this work, however, the
simulations should start with a system not having the
predetermined mass density profile. That is to say, spurious
"evolution" of the system initialized with the Maxwellian
approximation still cannot be avoided in their way.

Here we would like to further demonstrate the significance of
initializing a system with the distribution function by exploring
the dynamical evolution of a compact super star cluster (SSC
hereafter) in a DM dominated galaxy, depicted with the Burkert
profile. The term "compact" here means no tidal stripping is
effected. In this case, the most important process involved is the
dynamical friction the compact SSC experiences, which reads as
(\cite{bin87})

\begin{equation}
  \frac{d\mathbf{V_{M}}}{dt}=-16\pi^{2}\ln\Lambda G^{2}m(M+m)\frac{\int_{0}^{V_{M}}f(v_{m})v_{m}^{2}dv_{m}}{V_{M}^{3}}\mathbf{V_{M}}
\end{equation}
where
\begin{equation}
  \Lambda=\frac{b_{max}V_{typ}^{2}}{G(M+m)}
\end{equation}

\noindent M and $V_{M}$ (with $V_{M}=|\mathbf{V_{M}}|$) are,
respectively, the mass and velocity of the SSC experiencing the
dynamical friction. The mass of the DM particle, $m$, is much
smaller than M. The quantity $b_{max}$ is the maximum impact
parameter and $V_{typ}$ a typical speed. Neither $b_{max}$ nor
$V_{typ}$ is precisely defined. Following Binney \& Tremaine
(1987), we take $b_{max}\equiv2$ kpc and $V_{typ}\equiv V_{M}$.

If the DM has a Maxwellian velocity distribution with dispersion
$\sigma_{bkgd}$, the dynamical friction can be written as:
\begin{equation}
  \frac{d\mathbf{V_{M}}}{dt}=-\frac{2\pi\log(1+\Lambda^{2})G^{2}M\rho}{V_{M}^{3}}[erf(X)-\frac{2X}{\sqrt{\pi}}\exp
  (-X^{2})]\mathbf{V_{M}}
\end{equation}
where erf is the error function, and
\begin{equation}
  X\equiv\frac{V_{M}}{\sqrt{2}\sigma_{bkgd}}
\end{equation}
The velocity dispersion $\sigma_{bkgd}$ can be evaluated from the
Jeans equation.

\begin{figure}
   \begin{center}\vspace*{-36mm}
   \FigureFile(10.8cm,8.8cm){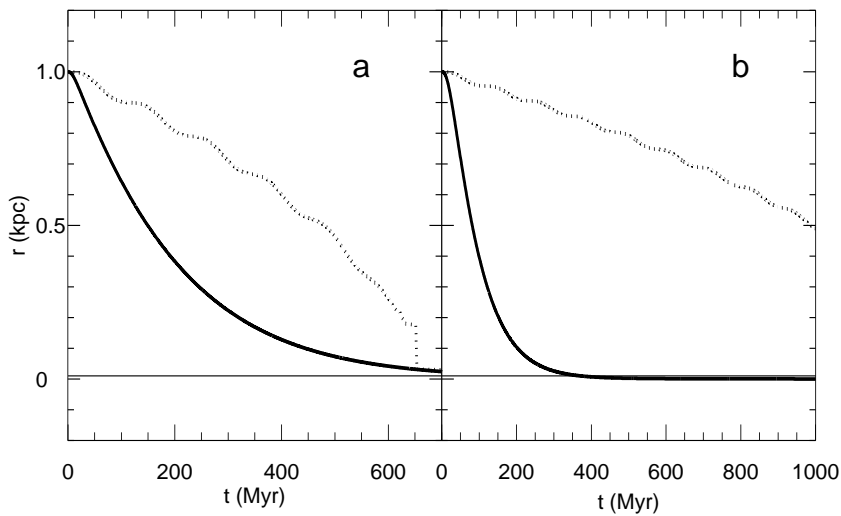}
   \end{center}\vspace*{-36mm}
      \caption{Sinking history of a compact SSC in the Burkert halo.
      Panel a) and b) present the situation for SSC with mass of $2 \times 10^{6} M_{\odot}$
      and $8 \times 10^{5} M_{\odot}$, respectively.
      The thick solid lines in both panels correspond to the case in Model A, i.e. the
      Burkert halo initialized with the
     distribution function, while the thick dotted ones in Model B: the Maxwellian halo. The
     thin solid line denotes the distance of 10 $pc$ to the center}
         \label{TrajTo}
   \end{figure}

Eq.(16) and eq.(18) are, respectively, the basis for considering
the dynamical friction in the DF Burkert halo and the Maxwellian
Burkert one. The initial SSC is assumed to move at a radius r=1kpc
with the local circular velocity. The sinking history of the SSC
is presented in Figure~\ref{TrajTo} a) and b) for SSC with
different masses. The thick solid lines in both panels express the
case in Model A, i.e. in the Burkert halo initialized using the
distribution function, while the thick dotted lines the case in
Model B, i.e. in the Maxwellian Burkert halo.

We can see from Figure~\ref{TrajTo}a) that the SSC with typical
mass of $2 \times 10^{6} M_{\odot}$ (\cite{fu03a}) sinks quickly,
in about $4 \times 10^{8} yr$, to the central region within a
hundred $pc$ in Model A, while takes much longer time in Model B.
It means that the SSC in Model A experiences much stronger
dynamical friction and loses angular momentum quickly at first.
This situation is consistent with the velocity distributions that
we discussed above. In fact, in an isotropic halo, the effect of
the dynamical friction comes from the DM particles whose
velocities are less than that of the SSC (\cite{bin87}). As we
illustrated in Figure~\ref{VelDisvsDisCen}, the true velocity
structure in the inner part of the halo is more peaked than a
Gaussian. Thus more particles with low velocities exist, resulting
in stronger dynamical friction on SSC. Inside 200 $pc$ from the
center, the SSC in Model A sinks slowly than that in Model B,
mainly due to its low velocity and low deceleration accordingly.

A similar sinking story is presented in Figure~\ref{TrajTo}b) for
a SSC with mass of $8 \times 10^{5} M_{\odot}$, where we can find
same trend of faster sinking in Model A than in Model B. The
interesting thing is that the SSC with smaller mass reaches the
center, less than 10 $pc$, in shorter period of time, as compared
with the corresponding situation in Model A shown in
Figure~\ref{TrajTo}a). The compact SSC with larger mass
experiences stronger dynamical friction, resulting in stronger
deceleration but longer sinking time.

As we know, when the dynamical friction exists the periodic motion
of a test particle in a spherical potential is damped like that of a
damped simple harmonic oscillator. If the dynamical friction is
strong enough, it is possible for the motion to lose completely its
oscillation behavior, the so-called over-damping. For an over-damped
SSC, the energy of SSC is in the form of potential energy in most of
time, which makes the energy loss by dynamical friction much less
effective than otherwise. For an incompact SSC, the tidal stripping
must not be ignored. In that case, the peeling process will make a
heavy SSC lighter, and the over-damping may not appear.

The shorter sinking time for SSCs in a DM halo initialized with
the steady-state distribution function should lead to higher
probability of forming massive bulges in late-type galaxies
(\cite{fu03a}), as well as of forming nuclear star clusters at
young ages (\cite{hua03}). It will also bring the intermediate
mass black holes, the seed black holes, within SSCs to the center
of a DM halo at early stage (\cite{fu03b}). All of these matter
are hot topics in astrophysics, which emphasizes the important
implications of using the distribution function to initialize the
system.

\section{CONCLUSION}
Constructing appropriate and reasonable initial conditions for an
isolated equilibrium system is one of the most important points to
various numerical simulations related to the formation and
evolution of galaxies. In this paper, we extend KMM's study on
cuspy density profiles to the Burkert profile. Using a time-saving
Monte Carlo method, we have shown clearly that the Burkert halo
initialized with the local Maxwellian approximation tends to
steepen, which is closely related to adopting a Gaussian as the
velocity distribution at any given points. This spurious evolution
leads to underestimating the dynamical friction on SSCs moving in
dark matter dominated galaxies. This important demonstration gives
us a valuable clue to clarify the embarrassed situation in our
previous investigation (\cite{fu03a}; \cite{hua03}), i.e. the
adopted Maxwellian approximation in these works causes longer
sinking time for SSCs so as not to form massive bulges in a few
Gyrs and not to form a young nuclear star cluster in the Burkert
halo. To initialize the dark matter halos with the steady-state
distribution function should be critical to what we are pursuing,
as what we have shown in this paper.

%\begin{acknowledgements}
%\end{acknowledgements}
\noindent  The authors would like to thank the referee, Andi
Burkert, for his valuable comments which improve the description of
the paper. This work is supported by NKBRSF G19990754 and NSFC
10373008. Fu is partly supported by the NSFC 10233020.

%%%
% See the manual for the detail.
%%%

\end{document}